# A Novel Video Compression Approach Based on Underdetermined Blind Source Separation

Jing Liu, Fei Qiao*, Qi Wei, Huazhong Yang
Department of Electronic Engineering
Tsinghua University
Beijing, China
Email: qiaofei@tsinghua.edu.cn

*Abstract*—This paper develops a new video compression approach based on underdetermined blind source separation. Underdetermined blind source separation, which can be used to efficiently enhance the video compression ratio, is combined with various off-the-shelf codecs in this paper. Combining with MPEG-2, video compression ratio could be improved slightly more than 33%. As for combing with H.264, 4X~12X more compression ratio could be achieved with acceptable PSNR, according to different kinds of video sequences.

## I. INTRODUCTION

Digital video is famous for its abundant information and vividness because of its rigid demand for the bandwidth and process power. Thanks to video coding technologies, various applications based on digital video become realizable. However, with the emergence of numerous such applications, bandwidth resource cannot meet the requirement of current needs any more. Therefore, high compression ratio is highly desired in video processing.

Blind Source Separation (BSS) provides a solution to recover original signals from several mixed signals. Independent Component Analysis (ICA) was widely accepted as a powerful solution of BSS since the past 20 years [1]. In 1999, A. Hyvarinen presented an improved ICA algorithm, called FastICA [2]. An detail overview on lots of algorithms on BSS is made and their usages on image processing are presented as well [3]. However, few researchers focused on utilizing BSS into video processing.

In this paper, we apply underdetermined BSS (meaning the number of original signals is more than that of mixed signals) to compress video sequences. A new codec defined as underdetermined blind source separation based video compression (UBSSVC) is developed. As we explained later in detail, UBSSVC has good performance on video compression.

This paper is organized as follows. The next section briefly reviews BSS problem. In section III, detailed structure of UBSSVC is stated. And then section IV shows the mixed video frames separation simulation results and the compression ratio improvements of UBSSVC. Finally, section V summarizes the superior and deficiency of this video compression method. Also, future work is proposed in this section.

## II. BLIND SOURCE SEPARATION AND SOLUTION TO UNDERDETERMINED BSS

Blind Source Separation was first established by J. Herault and C. Jutten in 1985 [1]. It can be described as following: multiple signals from separate sources $s(t) = [s_1(t), s_2(t), \ldots, s_n(t)]$ are somehow mixed into several other signals, defined as mixed signals or observed signals $x(t) = [x_1(t), x_2(t), \ldots, x_m(t)]$. The objective of BSS is to find an inverse system to get the estimation of original source signals. The output of the inverse system is estimated signals. It can be represented by $y(t) = [y_1(t), y_2(t), \ldots, y_n(t)]$. The reason for the "Blind" here is neither the source signals nor the mixed process are known to the observer.

Multiple researches were conducted on BSS the past decades. But most of them are investigated under the premise of $n < m$ and $n = m$. For $n < m$, BSS can be defined as overdetermined BSS, while for $n = m$, it can be defined as standard BSS. Here $n$ represents the number of source signals and $m$ represents the number of mixed signals. Independent Component Analysis (ICA) is the main solution for overdetermined and standard case. However, it is not suitable for $n > m$. Other methods like Spare Component Analysis (SCA) [4-7] and overcomplete ICA [8, 9], are investigated for the underdetermined case recent years. In fact, underdetermined BSS is still a tough challenge today.

### A. Overview of Blind Source Separation (BSS)

Let $s(t) = [s_1(t), s_2(t), \ldots, s_n(t)]$ be an $n \times T$ matrix, here $T$ is the sample number of each source signal. The $s_i, i = 1, 2, \ldots, n$ represent $n$ unknown original source signals. Suppose that there are $m$ ( $m \geq n$ ) sensors, the output of each sensor is denoted as $x_i, i=1,2,\ldots,m$, which is a linear combination of the $n$ source signals. The mixing model can be expressed as:

$$x = As; \quad A \in R^{m \times n}, s \in R^{n \times T} \qquad (1)$$

where $A$ is an $m \times n$ mixing matrix. $x(t) = [x_1(t), x_2(t), \ldots, x_m(t)]$ represents the mixed signals. Both $A$ and $s$ are unknown, while $x$ is known to observer.

The purpose of BSS is to recover $s$ only based on mixed signals $x(t) = [x_1(t), x_2(t), \ldots, x_m(t)]$ and sometimes prior knowledge of source signals $s(t) = [s_1(t), s_2(t), \ldots, s_n(t)]$. The separation model is shown as following:

$$y = Wx, \quad W \in R^{n \times m}, \quad x \in R^{m \times T} \qquad (2)$$

where $W$ is an $n \times m$ separating matrix. $y(t) = [y_1(t), y_2(t), \ldots, y_n(t)]$ is the estimation of source signals $s(t) = [s_1(t), s_2(t), \ldots, s_n(t)]$ and is named after estimated signals.

Combining (1) and (2), the following equation can be gotten:

$$y = Wx = WAs = Cs, \quad C \in R^{n \times n} \qquad (3)$$

where $C = WA$ is an $n \times n$ matrix. If $C = I$, where $I$ is the identity matrix, the source signals are perfectly recovered. However, it is not always the real case. For most applications, the amplitude and the order of source signals have little influence on their contained information. Therefore, $C = P$ is acceptable, where $P$ is the permutation matrix. The basic framework of BSS is shown in figure 1 [3].

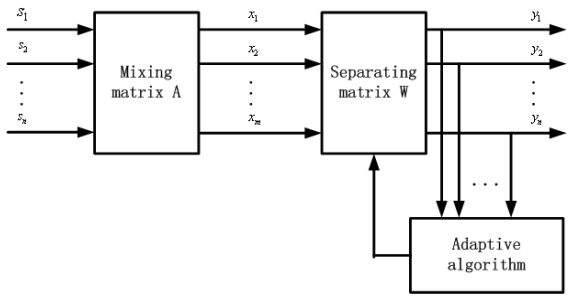

Figure 1. Basic framework of BSS

### B. Underdetermined Blind Source Separation

The above solution is only suitable for the scenario when $n \leq m$. That's because for the underdetermined BSS, separating matrix $W$, which meets $C = WA = P$, does not exist. Therefore, solutions for underdetermined BSS are different from the $n \leq m$ case. One typical method, Sparse Component Analysis (SCA), could be used to efficiently and effectively solve the underdetermined BSS problem.

SCA uses the sparsity of source signals to compensate information loss in the mixing process. So specific assumptions of mixing matrix $A$ and source matrix $s$ should be considered as follows [4]:

1) Any $m \times m$ square sub-matrix of mixing matrix $A \in R^{m \times n}$ is nonsingular;

2) There are at most $m-1$ nonzero elements of any column of matrix $s$.

If the above assumptions are satisfied, the source matrix $s$ can be recovered by SCA.

### III. PROPOSED UBSSVC METHOD

The proposed UBSSVC framework combines UBSS and conventional codec together, which is shown in Figure 2. In mixing process, $n$ frames, $f_{i,1}, f_{i,2}, \cdots, f_{i,n}$, are mixed into $m$ frames, $\tilde{f}_{i,1}, \tilde{f}_{i,2}, \cdots, \tilde{f}_{i,m}$. In separating process, $m$ frames, $\tilde{f}'_{i,1}, \tilde{f}'_{i,2}, \cdots, \tilde{f}'_{i,m}$, are separated to $n$ frames, $f'_{i,1}, f'_{i,2}, \cdots, f'_{i,n}$. The encoding process and decoding process are discussed in detail as follows.

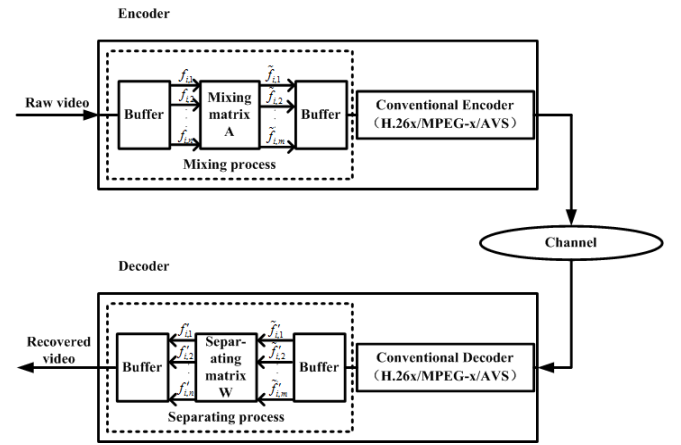

Figure 2. Framework of UBSSVC

### A. Encoder Structure

The mixing process of UBSS is applied to compress video before the conventional encoder. Consider a video sequence with $b \times n$ frames, $s_1, s_2, \cdots, s_{b \times n}$; where $s_i \in R^T$ is a T-pixel frame. The encoder first chooses a matrix $A \in R^{m \times n} (m < n)$ to mix these frames. Thus, the number of frames is decreased to $b \times m$. And then these $b \times m$ mixed frames are encoded by traditional encoder such as MPEG-2, H.264.

Unlike the traditional scenario of the underdetermined BSS issue, mixing process is factitious in this proposed method. Thus, a specific mixing matrix $A$, known by both encoder and decoder, is used to mix raw video frames.

As explained above, information loss happens in mixing process of underdetermined BSS. So when choosing the mixing matrix $A$, efforts should be made to decrease the information loss. Thus, in different mixed frames, the weight of different original frames should be varied. As each component of a row of $A$ can be treated as the weight of every original frame in a mixed frame, the components of a row of $A$ should be varied largely from each other.

### B. Decoder Structure

In the proposed decoder structure, received data is firstly decoded by traditional decoder; then the source recovery algorithm of underdetermined BSS is applied to recover original video sequence.

As the matrix $A$ is known exactly by decoder, the order of separated frames is the same with that of original frames. The frames' order of recovered video sequence is not disturbed by mixing process and separating process, which is different from traditional BSS.

The key technology in this decoding process is the recovery algorithm. In this work, SCA is adopted to separate mixed frames.

Let $x_i, i=1,2,\ldots,m$ and $s_i, i=1,2,\ldots,n$ represent mixed frames and source frames respectively; and $a_j, j=1,\cdots,n$ is the $j-th$ column of mixing matrix $A$. Therefore, the mixing process at encoder side can be described as following.

$$x(t) = \begin{pmatrix} x_1(t) \\ x_2(t) \\ \vdots \\ x_m(t) \end{pmatrix} = a_1 s_1(t) + a_2 s_2(t) + \cdots + a_n s_n(t) \quad (4)$$

Given the mixing matrix $A$ satisfies the assumption 1), any $m-1$ columns of $A$ span a m-dimensional linear hyperplane $\mathcal{H}_q$, which can be denoted as $\mathcal{H}_q = \{h \mid h \in R^m, \lambda_{ik} \in R, h = \lambda_{i_1} a_{i_1} + \cdots + \lambda_{i_{m-1}} a_{i_{m-1}}\}$, where $q = 1,\cdots,C_n^{m-1}$. If source matrix $s$ satisfies assumption 2), it is reasonable to suppose that at $t$ moment, all source frames except for $s_{i_1}, s_{i_2}, \cdots, s_{i_{m-1}}$ are zero, where $\{i_1, i_2, \cdots, i_{m-1}\} \subset \{1, 2, \cdots, n\}$. Consequently, at $t$ moment, (4) can be rewritten as

$$\begin{pmatrix} x_1(t) \\ x_2(t) \\ \vdots \\ x_m(t) \end{pmatrix} = a_{i_1} s_{i_1}(t) + a_{i_2} s_{i_2}(t) + \cdots + a_{i_{m-1}} s_{i_{m-1}}(t) \quad (5)$$

From (5), it can be concluded that the column vector of observed signals matrix $x$ at $t$ moment is in one of $C_n^{m-1}$ hyperplanes $\mathcal{H}$. Therefore, at decoder side, mixed frames can be recovered by the following algorithm.

1) Get the set $\mathcal{H}$ of $C_n^{m-1}$ m-dimensional hyperplanes which are spanned by any $m-1$ columns of $A$;

2) Repeat for $j=1:m$

   i) If $x_j = x(:,j)$ is in a hyperplane $\mathcal{H}_q$, then the following equation can be gotten

$$x_j = \sum_{k=1}^{m-1} \lambda_{i_k, j} a_{i_k} \quad (6)$$

   ii) Comparing equation (5) and (6), $s_i = s(:,i)$ can be recovered: its components are $\lambda_{i_k, j}$ in the place $i_k, k=1,\cdots,m-1$, and other components equal to zero.

## IV. EXPERIMENT RESULTS

In order to validate this approach, multiple simulations are performed on four standard test video sequences: hall, container, foreman and football. The football sequence has the largest temporal variations, followed by foreman, and container ranks the third, while the hall sequence contains the most slowly scene variations. Fig. 3 shows the first frames of these four test sequences. In our experiments, the first 40 frames of each test sequence are used. Peak-Signal-to-Noise Ratio (PSNR) is used to evaluate the performance of recovery algorithm. PSNR is calculated by the following formulas, where $\widehat{s}_i$ is estimation of $s_i$, $M$ and $N$ is the width and height of frames.

$$PSNR = 10\log_{10} \frac{255 \times 255}{MSE} \quad (7)$$

$$MSE = \frac{\sum_{i=1}^{n}(\widehat{s}_i - s_i)^2}{M \times N} \quad (8)$$

In our experiments, the mixing matrix $A \in R^{3\times 4}$, shown in (9), is chosen to mix raw video sequence. The mixing process is performed as follows: continuous 4 frames are taken as source signals $s$, then $A$ multiplies by $s$ to calculate the mixed frames $x$. 30 mixed frames are generated after the mixing process. And then the above algorithm is applied to separate these mixed frames. To satisfy the requirement for sparsity, mixed frames are first transformed by a 2-D discrete Haar wavelet transform. SCA is used to recover the sparse high frequency components, while the recovered low frequency components are equal to multiply generalized inverse of mixing matrix $A$ by mixed low frequency components. Fig. 4 shows the recovery results on different video test sequences. Although for some sequence, such as football, PSNR is a little low, it is still enough for monitor applications with low resolution requirements.

$$A = \begin{pmatrix} 0.50 & 0.75 & 0.25 & 0.15 \\ 0.40 & 0.25 & 0.10 & 1.00 \\ 0.45 & 0.10 & 0.85 & 0.25 \end{pmatrix} \quad (9)$$

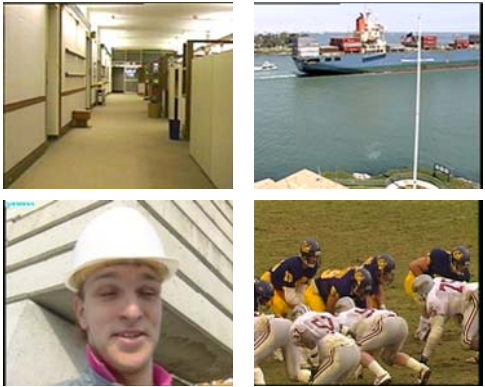

Figure 3. The first frames of four test sequences

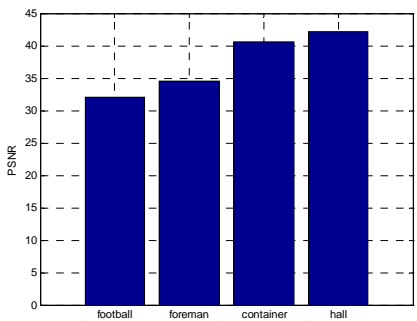

Figure 4. Separating results on four test sequences

The compression results for MPEG-2, UBSSVC+MPEG-2, H.264 and UBSSVC+H.264 are shown in table I. From the results, UBSSVC indeed can gain higher compression ratio. Combining with MPEG-2, video compression ratio could be improved slightly more than 33%. As for combing with H.264, 4X~12X more compression ratio could be achieved with acceptable PSNR, according to different kinds of video sequences.

TABLE I. RESULT OF VIDEO ENCODED BY MPEG-2

|  | hall | container | foreman | football |
|---|---|---|---|---|
| **Original size(MB)** | 1.45 | 1.45 | 1.45 | 1.45 |
| **MPEG-2(KB)** | 225 | 225 | 225 | 225 |
| **UBSSVC+MPEG-2(KB)** | 169 | 169 | 169 | 169 |
| **H.264(KB)** | 11.6 | 10.1 | 18.9 | 98.2 |
| **UBSSVC+H.264(KB)** | 3.89 | 3.15 | 3.80 | 7.31 |

## V. CONCLUSION

This paper initially develops the novel video compression approach UBSSVC. Furthermore, experiments are conducted to validate the efficiency of recovery algorithm and to measure the video compression ratio improvements of UBSSVC.

The proposed method is suitable for video surveillance system perfectly. Firstly, it can achieve higher video compression ratio to decrease the bandwidth resource utilization . Secondly, the computation complexity of mixing process at encoder side is low, which may not largely increase the power consumption, when improving the video compression ratio. What's more, the mixing and separating process of UBSS has great potential in low-complexity video compression. However, the presented new method still has more issues to be improved in our future work. Like how to improve the compression ratio gained by the mixing process and enhance the separating results of video quality.

References

[1] A. Hyvarinen, J. Karhunen, and E. Oja, *Independent component analysis*. New York: J. Wiley, 2001.
[2] A. Hyvarinen, "Fast and robust fixed-point algorithms for independent component analysis," *Neural Networks, IEEE Transactions on,* vol. 10, pp. 626-634, 1999.
[3] A. Cichocki and S. i. Amari, *Adaptive blind signal and image processing : learning algorithms and applications*. Chichester, England ; New York: John Wiley, 2002.
[4] P. Georgiev, F. Theis, and A. Cichocki, "Sparse component analysis and blind source separation of underdetermined mixtures," *Neural Networks, IEEE Transactions on,* vol. 16, pp. 992-996, 2005.
[5] Y. Q. Li, A. Cichocki, S. I. Amari, S. Shishkin, J. T. Cao, and F. J. Gu, "Sparse representation and its applications in blind source separation," in *Advances in Neural Information Processing Systems 16*. vol. 16, S. Thrun, K. Saul, and B. Scholkopf, Eds., ed Cambridge: M I T Press, 2004, pp. 241-248.
[6] M.-r. Ren and P. Wang, "Underdetermined Blind Source Separation Based on Sparse Component," in *Electronic Computer Technology, 2009 International Conference on*, 2009, pp. 174-177.
[7] S. Zhenwei, T. Huanwen, and T. Yiyuan, "Blind source separation of more sources than mixtures using sparse mixture models," *Pattern Recognition Letters,* vol. 26, pp. 2491-24992499, Dec. 2005.
[8] T. W. Lee, M. S. Lewicki, M. Girolami, and T. J. Sejnowski, "Blind source separation of more sources than mixtures using overcomplete representations," *Ieee Signal Processing Letters,* vol. 6, pp. 87-90, Apr 1999.
[9] K. Waheed, F. M. Salem, and Ieee, "Algebraic independent component analysis: An approach for separation of overcomplete speech mixtures," in *Proceedings of the International Joint Conference on Neural Networks 2003, Vols 1-4*, ed New York: Ieee, 2003, pp. 775-780.